\documentclass[11pt,a4paper]{article}
\newcommand{\captionfonts}{\small}

\makeatletter  
\long\def\@makecaption#1#2{%
  \vskip\abovecaptionskip
  \sbox\@tempboxa{{\captionfonts #1: #2}}%
  \ifdim \wd\@tempboxa >\hsize
    {\captionfonts #1: #2\par}
  \else
    \hbox to\hsize{\hfil\box\@tempboxa\hfil}%
  \fi
  \vskip\belowcaptionskip}
\makeatother   

\usepackage{amssymb}
\usepackage{graphicx}
\usepackage{hyperref}

\title{The WITCH experiment: Acquiring the first recoil ion spectrum\thanks{Proceedings of EMIS07, published NIM B}}
\date{}

\begin{document}


\maketitle
\begin{center}
V.Yu.~Kozlov \footnotemark$\;^{a)}$, M.~Beck$\,^{b)}$, 
S.~Coeck$\,^{a)}$, P.~Delahaye$\,^{c)}$, P.~Friedag$\,^{b)}$, M.~Herbane$\,^{a)}$, A.~Herlert$\,^{c)}$,
I.S.~Kraev$\,^{a)}$, M.~Tandecki$\,^{a)}$, S.~Van~Gorp$\,^{a)}$, F.~Wauters$\,^{a)}$, Ch.~Weinheimer$\,^{b)}$, 
F.~Wenander$\,^{c)}$, D.~Z\'{a}kouck\'{y}$\,^{d)}$ and N.~Severijns$\,^{a)}$
\end{center}

\begin{center}
\noindent $^{a)}$ {\small \it K.U.Leuven, Instituut voor Kern- en Stralingsfysica, Celestijnenlaan~200D, B-3001~Leuven, Belgium}
\newline
\noindent $^{b)}$ {\small \it Universit\"{a}t M\"{u}nster, Institut f\"{u}r Kernphysik, Wilhelm-Klemm-Str. 9, D-48149~M\"{u}nster, Germany}
\newline
\noindent $^{c)}$ {\small \it CERN, CH-1211~Gen\`eve 23, Switzerland}
\newline
\noindent $^{d)}$ {\small \it Nuclear Physics Institute, ASCR, 250 68 \v{R}e\v{z}, Czech Republic}
\end{center}

\begin{abstract}
\small
The standard model of the electroweak interaction describes 
$\beta$-decay in the well-known V-A form. Nevertheless, the most  
general Hamiltonian of a beta-decay includes also other possible  
interaction types, e.g. scalar (S) and tensor (T) contributions, which  
are not fully ruled out yet experimentally. The WITCH experiment aims  
to study a possible admixture of these exotic interaction types in  
nuclear $\beta$-decay by a precise measurement of the shape of the  
recoil ion energy spectrum. The experimental set-up couples a double  
Penning trap system and a retardation spectrometer. The  
set-up is installed in ISOLDE/CERN and was recently shown to be  
fully operational. The current status of the experiment is presented together with the data acquired during the 2006 campaign, showing the first recoil ion energy spectrum obtained. The data taking procedure and corresponding data acquisition system are described in more detail. Several further technical improvements are briefly reviewed.
\end{abstract}

\begin{flushleft}

\end{flushleft}
\footnotetext{Corresponding author, Valentin.Kozlov@gmail.com}

\section{Introduction}
\label{intro}
The weak interaction described by the Standard Model has the well-known V~(vector) $-$ A~(axial-vector) structure. The most general Hamiltonian for nuclear $\beta$-decay, however, suggests more possibilities consistent with Lorentz-invariance \cite{lee56, jackson57a}: scalar~(S), tensor~(T) and pseudoscalar~(P) interactions. The presence of S- and T- contributions in weak interaction is not yet fully ruled out experimentally, i.e. the present constraints are at the level of about 8\% (95\%~C.L.) of the V- and A-interaction \cite{severijns06}. One of the probes to search for these exotic interactions is to study the $\beta-\nu$ angular correlation. This correlation for unpolarized nuclei can be characterized by the $\beta-\nu$ angular correlation coefficient \emph{a} the value of which depends on the type of interaction involved and is for instance for pure Fermi transitions given by

\begin{eqnarray}
a_{F} & = &
\frac{\left|C_{V}\right|^{2}+\left|C_{V}^{\prime}\right|^{2}-\left|C_{S}\right|^{2}-\left|C_{S}^{\prime}\right|^{2}}{\left|C_{V}\right|^{2}+\left|C_{V}^{\prime}\right|^{2}+\left|C_{S}\right|^{2}+\left|C_{S}^{\prime}\right|^{2}}\,
\label{eq:aF}
\end{eqnarray}


From the properties of interactions it can be shown that this coefficient $a$ also determines the shape of the recoil ion energy spectrum \cite{beck03a}. The primary goal of the WITCH experiment is to measure this recoil ion energy spectrum with high enough precision in order to infer new  constraints on the S- or T- contributions. The issue of exotic scalar and tensor interactions was recently addressed in a number of experiments in nuclear physics \cite{adelberger99, scielzo04, gorelov05, rodriguez07, lienard07} as well as particle physics (for example \cite{frlez04}). One should also note here that the presence of these interaction types implies the existence of corresponding mediator bosons. An extensive review can be found in \cite{severijns06, herczeg01}.


\section{Experiment}
\label{exp}
\begin{figure}
	\centering
		\includegraphics[width=0.8\textwidth]{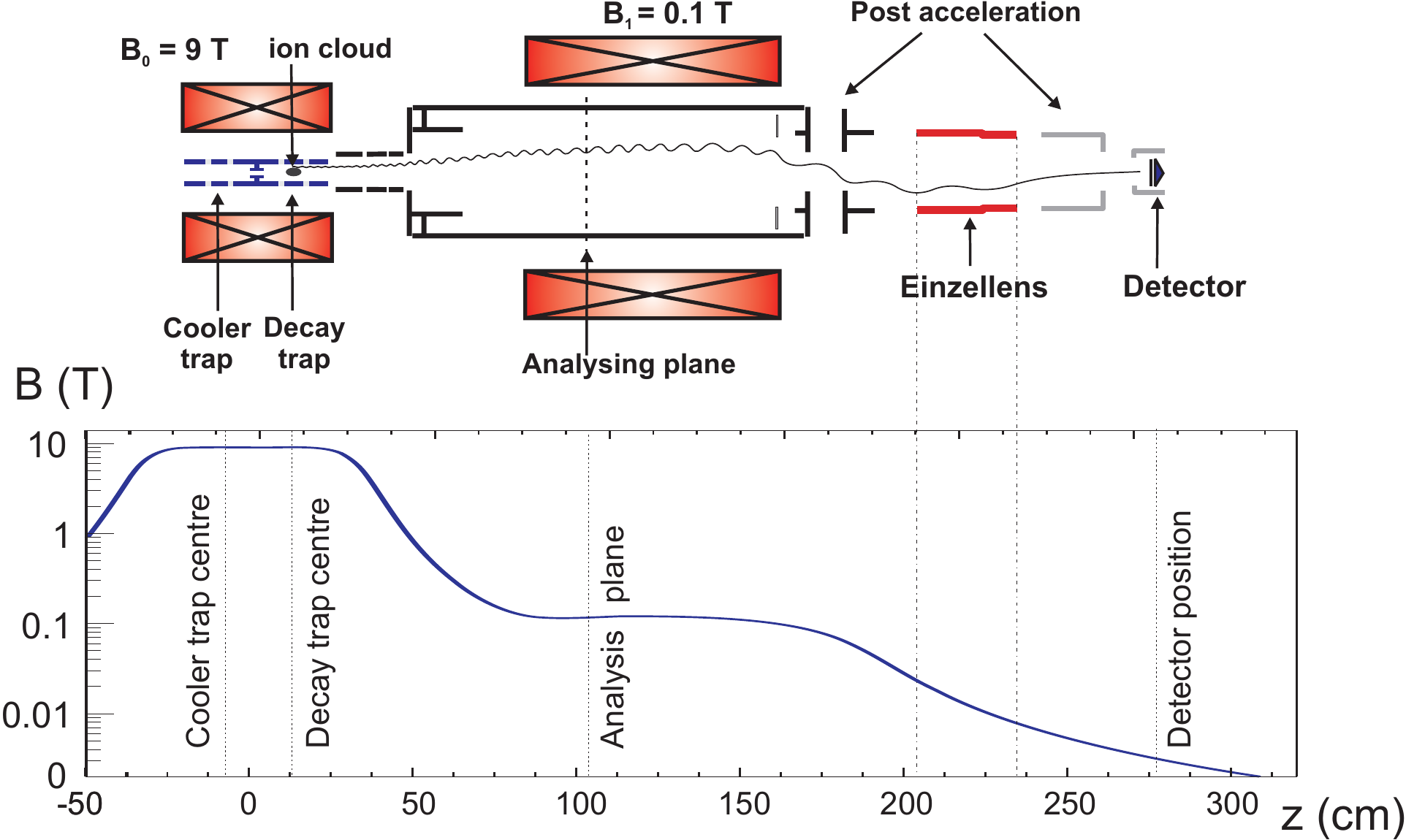}
	\caption{Principle of the WITCH experiment: the retardation spectrometer is shown together with the corresponding magnetic field profile. z=0 corresponds to the center of the 9 T magnet. The positions of the traps, the analysis plane and the recoil ion detector are indicated.}
	\label{fig:spectrometer_b-field}
\end{figure}

The WITCH set-up is installed at ISOLDE/CERN \cite{kugler00}, thus fulfilling the requirement of the experiment for a high intensity radioactive source. The quasi-continuous beam produced at ISOLDE is first trapped and bunched by the REXTRAP facility \cite{ames05}. The ion bunch ejected from REXTRAP is then guided into the WITCH set-up. The principle of the set-up is based on a combination of a double Penning trap system to form a scattering-free source and a retardation spectrometer (MAC-E filter) to measure the energy of the recoil daughter ions (Fig.\ref{fig:spectrometer_b-field}). This combination was chosen to obtain high accuracy in measuring recoil spectra with the end point energy of the order of 100~eV. First, the ions are decelerated with a pulsed drift tube \cite{coeck07a} in order to be loaded into the first Penning trap, called cooler trap. There they are cooled by buffer gas collisions and mass selectively purified. The ion cloud is then transferred into the second Penning trap, the decay trap. The energy of the ions leaving the decay trap after $\beta$-decay is probed by an electrostatic retardation potential. Both traps are placed in a 9~T magnetic field while the retardation analysis plane is at 0.1~T field. According to the working principle of the retardation spectrometer this leads to a 98.9\% conversion of the radial energy into axial energy at the analysis plane. The ions of charge state $q$ that pass the retardation potential are accelerated to $qU$~eV, with $U \sim 10$~kV, and focused towards a position sensitive micro-channel plate detector (MCP). Position sensitivity is realized by delay line anodes, i.e. the one-dimensional position of the particle hit is deduced from the difference of the propagation times to both ends of the corresponding wire~\cite{lienard05}. Varying the retardation potential over the necessary range allows to measure the full integral recoil ion spectrum. A more detailed description of the WITCH experiment can be found in \cite{beck03a,kozlov06}.

\section{Current status}
\label{status}
The main effort during the year 2006 was to improve the overall efficiency of the set-up \cite{coeck07a, coeck06} including optimization of REXTRAP for the needs of the WITCH experiment and to extend the data acquisition (DAQ) system to control more of the important parameters (sec.~\ref{daq}). An overview of the present efficiencies is given in Table~\ref{tab:eff}. There is good improvement in relation to year 2004~\cite{kozlov06} but one order of magnitude is still missing in comparison with the ideal set-up, mainly due to a not yet fully optimized injection into the magnetic field. Another recent achievement is the mass resolving power of the cooler trap $m/\Delta m\sim1\div2\cdot10^{5}$, thanks to newly produced traps that were silver- and gold-plated in GSI (Germany).

\begin{table}[!h]
\caption{Estimate of the presently best achieved total efficiency in comparison with the values for the ideal set-up (for parameters that were not yet studied the
values of an ideal set-up are taken to calculate the total
efficiency).}
\label{tab:eff}       
\centering
\vspace*{3mm}
\begin{tabular}{lcc}
\hline \rule{0pt}{10pt}\textbf{Description}& \textbf{ideal}&
\textbf{Best achieved}\tabularnewline & \textbf{set-up}&
\textbf{2004-2006}\tabularnewline \hline \rule{0pt}{10pt}Beamline
transfer + pulse down& 50\%& $\sim$80\%\tabularnewline Injection
into B-field& 100\%& $\sim$20\%\tabularnewline  Cooler trap efficiency&
100\%& $\sim$50\%\tabularnewline  Transfer between traps& 100\%&
$\sim$70\%\tabularnewline  Storage in the decay trap& 100\%&
100\%\tabularnewline  Fraction of ions leaving the decay trap&
$\sim$40\%& not studied\tabularnewline  Shake-off for charge state n=1&
$\sim$10\%& not studied\tabularnewline  Transmission through the
spectrometer& 100\%& 100\% (prelim.)\tabularnewline MCP efficiency&
60\%& 52.3(3)\% \cite{lienard05}\tabularnewline \hline
\rule{0pt}{10pt}\textbf{Total efficiency}& \textbf{$\sim$1\%}&
\textbf{$\sim$0.12$$\%}\tabularnewline
\end{tabular}
\end{table}

While improving the efficiency of the set-up several new problems showed up as well. The major ones are the discharges in the cooler trap and in the spectrometer, related to the buffer gas pressure and $\gamma$-activity present in the system, and a sparking of the acceleration part for voltages above 7~kV. This limited the functionality of the system but did not prevent us from measuring the first recoil ion spectrum.

\section{First recoil ion spectrum}
\label{spec}
\begin{figure}
	\centering
		\includegraphics[width=0.75\textwidth]{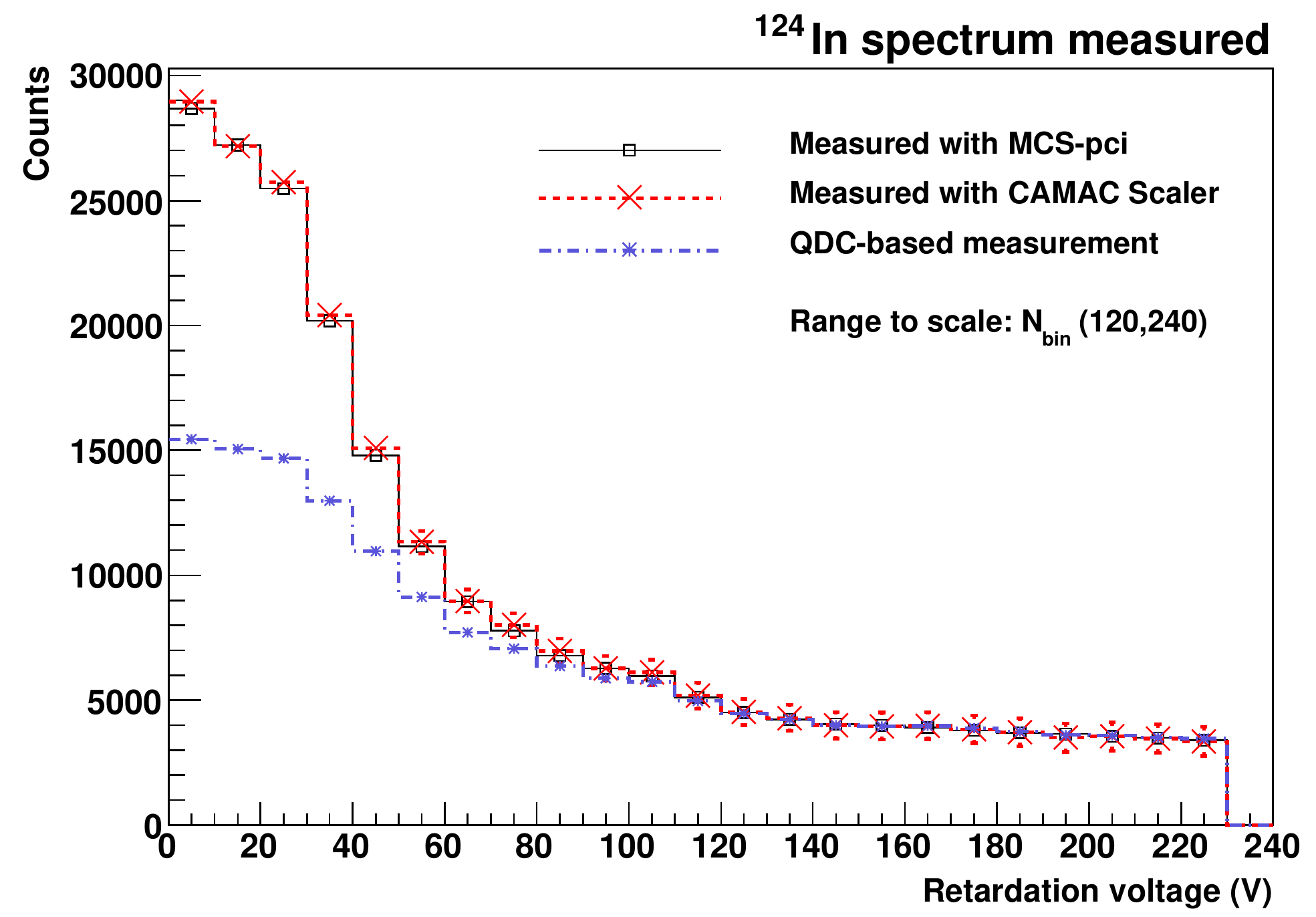}
	\caption[]{First recoil ion spectrum measured. The incoming beam is a mixture of $^{124g}$In and $^{124m}$In; 23 retardation steps in the range of 0$\div$220~V. Three spectra shown correspond to different methods of data acquisition (see sec.~\ref{daq}).}
	\label{fig:124In_meas}
\end{figure}

\begin{figure}
	\centering
		\includegraphics[width=0.75\textwidth]{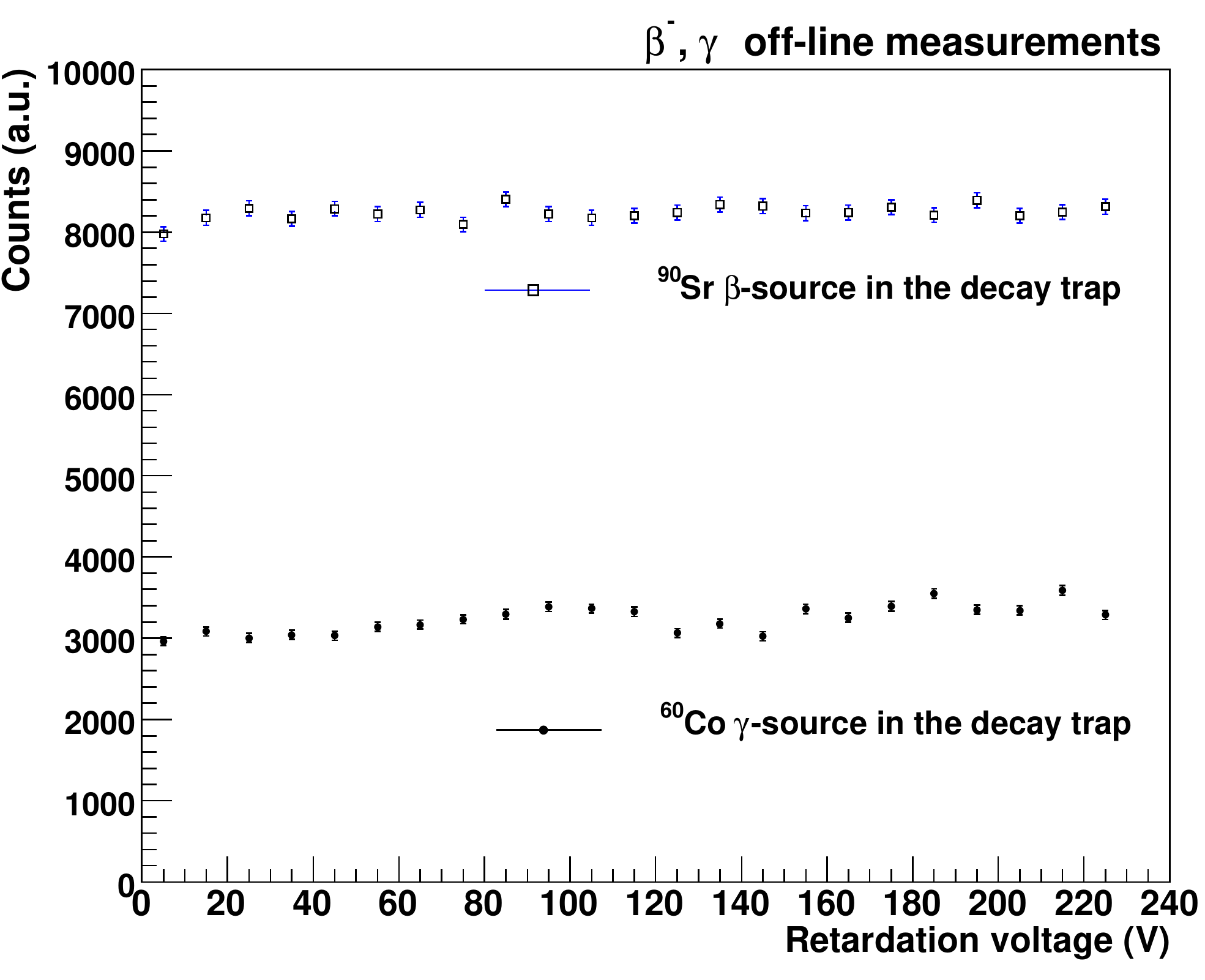}
	\caption{Off-line measurements with the $\beta$-source ($^{90}$Sr) (top) and $\gamma$-source ($^{60}$Co) (bottom) in the decay trap. Absolute intensities are not to be compared due to different sources strength and measurement time. The running cycle was exactly the same as during the on-line measurement of $^{124}$In.}
	\label{fig:offline_meas}
\end{figure}

To perform a proof-of-principle experiment a high beam intensity is required. Moreover, one can switch to a beta-minus isotope in which case all recoil daughter ions have a non-zero charge contrary to the beta-plus case where one has to rely on the shake-off probability to obtain 1+ ions because of the 1+ charge state of the incoming mother ions. In this way one can compensate for the missing order of magnitude in the overall set-up efficiency (Tab.~\ref{tab:eff}). Thus, $^{124}$In was chosen for the proof-of-principle experiment. A drawback of this isotope is the complicated decay scheme, with a significant amount of $\gamma$'s, and that the incoming beam from ISOLDE contains both ground state $^{124g}$In and isomer $^{124m}$In. The end point energy of the $^{124g}$In and $^{124m}$In recoil ions from $\beta$-decay is different, being 196~eV and 83~eV respectively. In addition, the recoil ion spectra of both states are influenced by the $\gamma$-radiation which follows the $\beta$-decay. This fact has to be correctly taken into account in the analysis~\cite{coeck07b}. The discharge problem in the spectrometer mentioned above did not allow to measure the recoil energy spectrum by using the retardation spectrometer. The Einzellens electrode (Fig.~\ref{fig:spectrometer_b-field}) was used instead. Discharges in the cooler trap were avoided by using a lower buffer gas pressure. First, to verify that recoil ions are indeed present in the system a simple Off/On cycle was executed, i.e. switching between no retardation and full retardation of the recoil ions. The expected significant drop in a count rate when the full retardation is applied was observed. Then the measurement cycle was changed to 23 retardation steps going up from  0~V to 220~V and the first recoil ion spectrum was obtained (Fig.~\ref{fig:124In_meas}). The data presented were acquired during $\sim$50~min. To verify that the measured spectrum originates from the recoil ions of $^{124}$In and is not caused by any correlated background, two off-line sources (a $\beta$-source of $^{90}$Sr and a $\gamma$-source of $^{60}$Co) were later installed for two independent measurements. The spectra measured under the same circumstances as during the on-line experiment are presented in Fig.~\ref{fig:offline_meas} showing no similarity to Fig.~\ref{fig:124In_meas}. The data taken with the $^{90}$Sr source together with GEANT4 simulations \cite{kozlov06} allow to estimate the $\beta$-background present during the on-line experiment. The fluctuation of data points taken with the $^{60}$Co source being larger than the statistical uncertainty suggests that certain discharges due to the $\gamma$-activity could still have been present. The detailed analysis of the measured $^{124}$In recoil spectrum is given in \cite{coeck07b}.

\section{Data acquisition}
\label{daq}
Two approaches are available for the recoil ion spectrum measurement: (1) fix the retardation value for each trap filling and (2) scan all retardation voltages during one trap filling. The first one is not directly possible at the moment since a normalization between different trap loads is necessary in this case (this is currently being developed, see sec.~\ref{outlook}) while the latter one requires that the half-life of the isotope is taken into account in the analysis. During the $^{124}$In measurement the second option was chosen, thus the recoil spectra in Fig.~\ref{fig:124In_meas} are not corrected for the half-lives of $^{124g}$In (3.11(10)~s) and $^{124m}$In (3.7(2)~s).

\begin{figure*}
	\centering
		\includegraphics[width=0.85\textwidth]{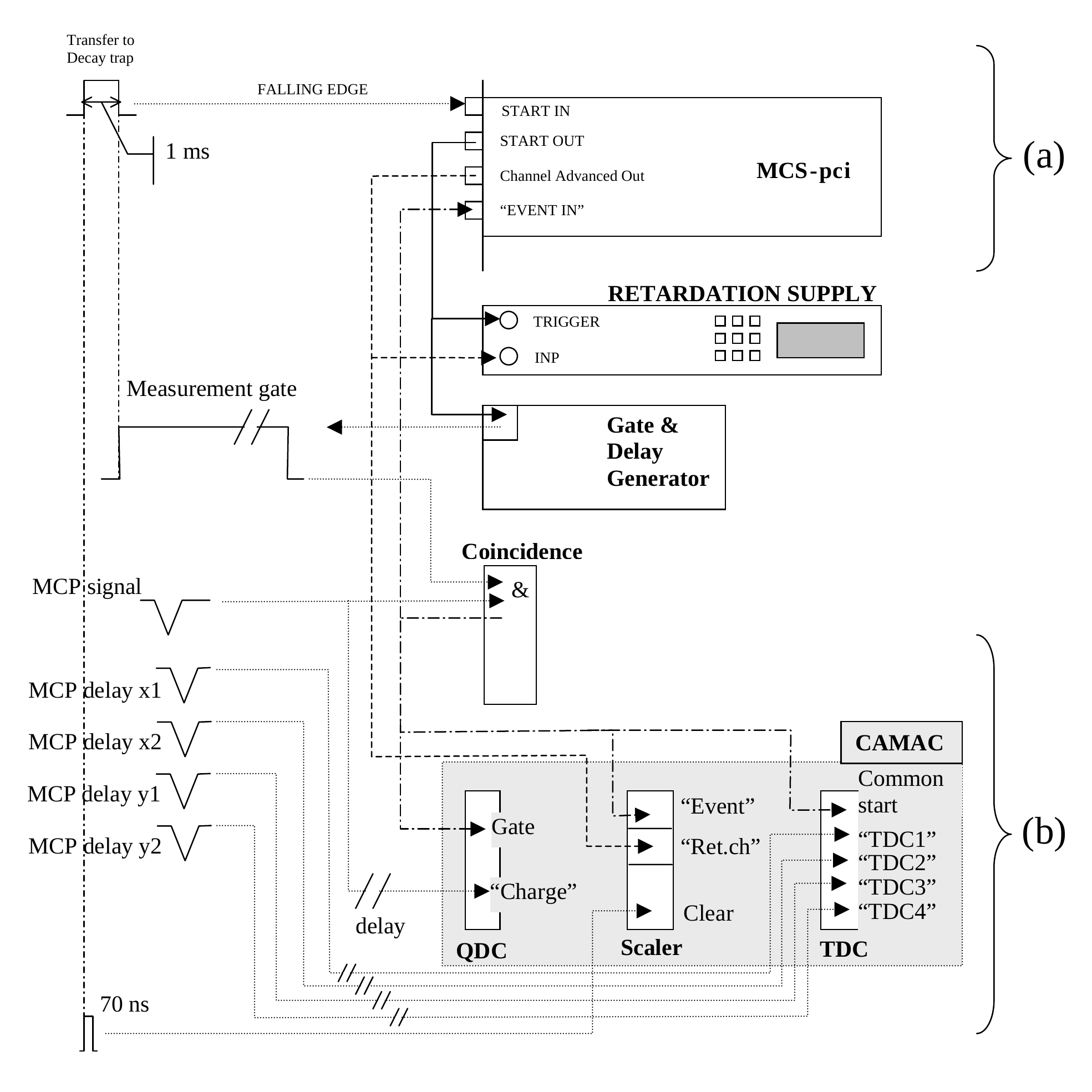}
	\caption{DAQ principle scheme of the WITCH experiment. Two branches of the system are indicated: (a) fast counting, based on the ORTEC MCS-pci card and (b) event-by-event, based on the CAMAC system.}
	\label{fig:daq-scheme}
\end{figure*}

\begin{figure}
	\centering
		\includegraphics[width=0.75\textwidth]{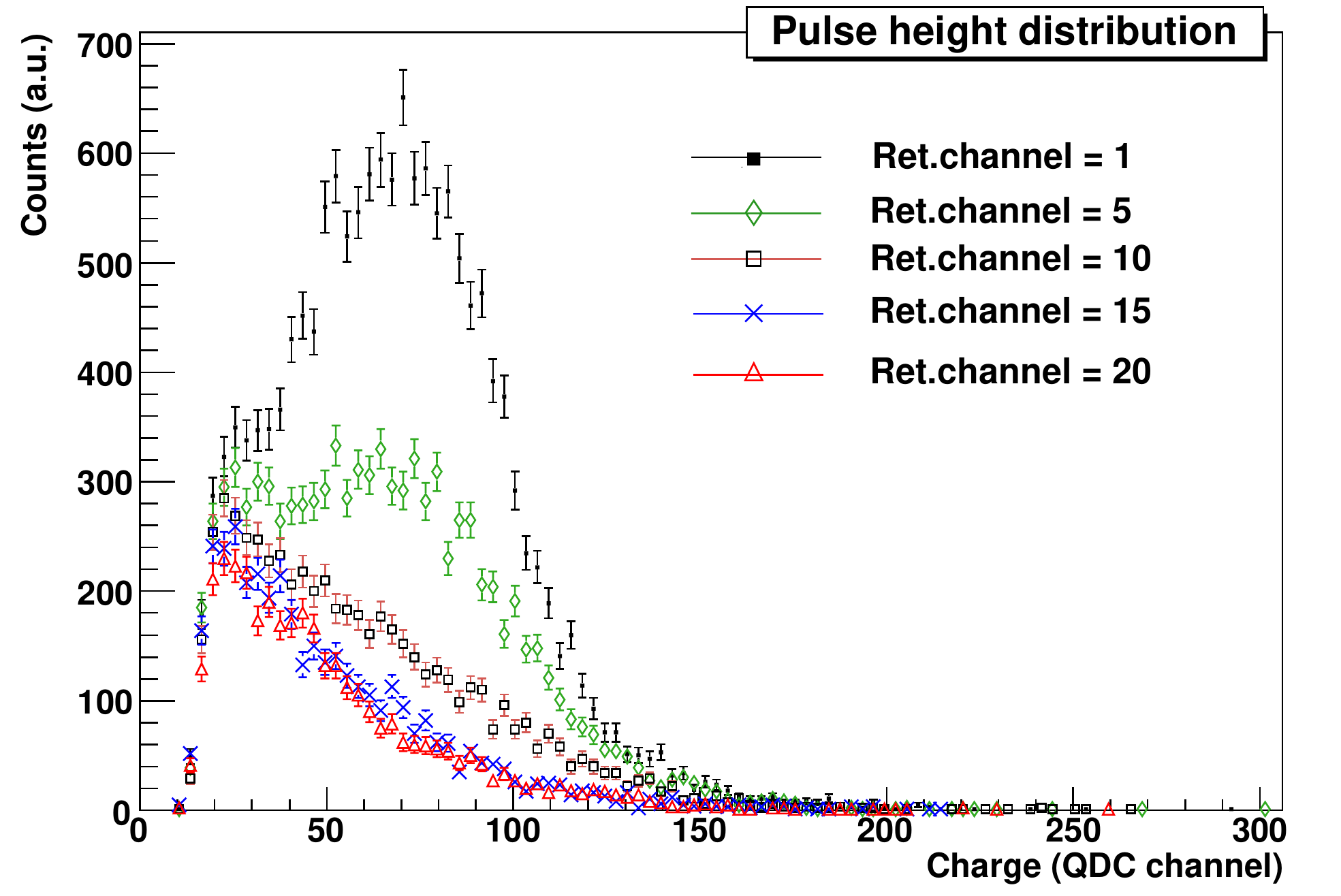}
	\caption{Pulse height distributions for different retardation steps.}
	\label{fig:phds_vs_retard}
\end{figure}

\begin{figure}
	\centering
		\includegraphics[width=0.75\textwidth]{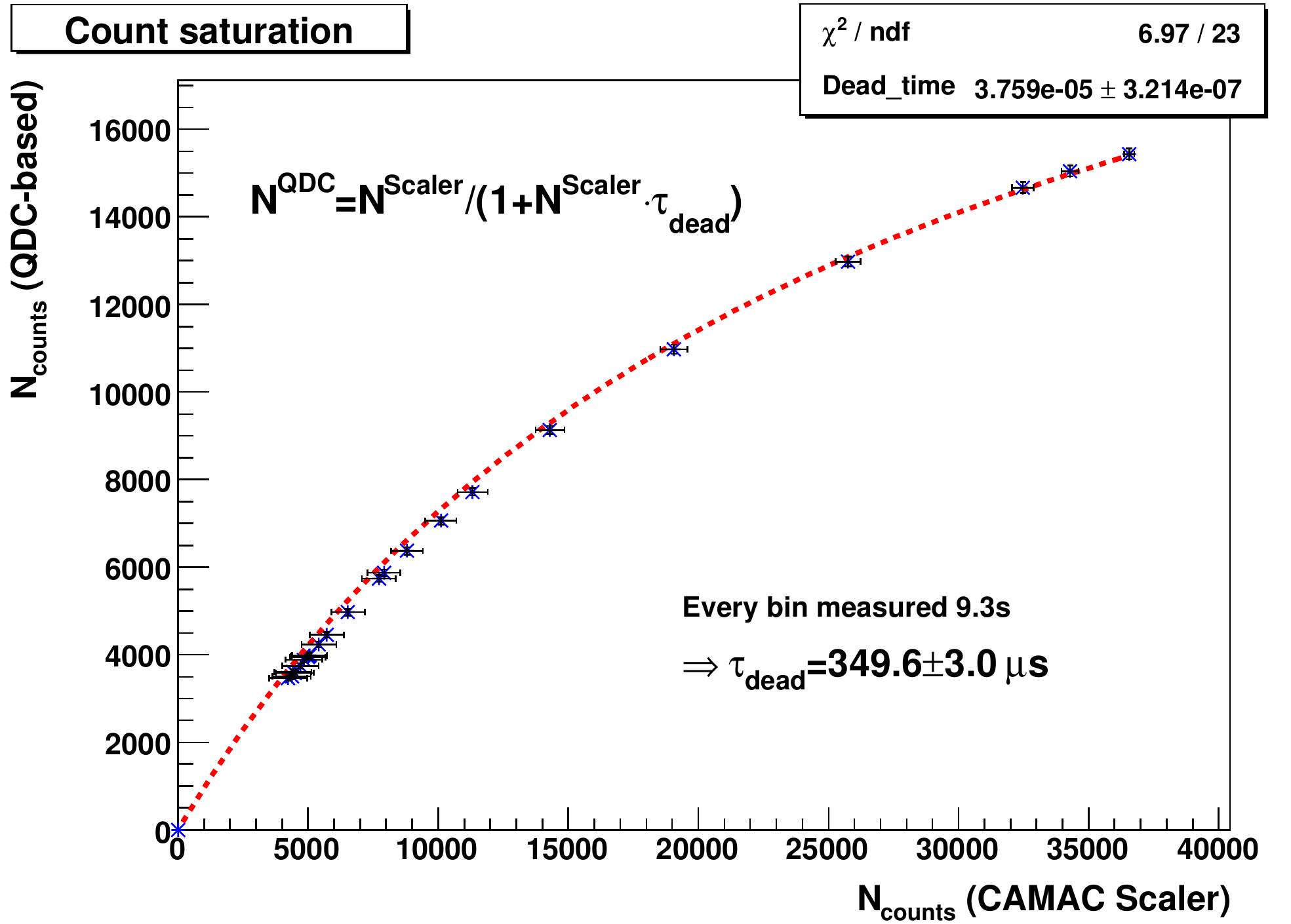}
	\caption{Saturation of the count rate in the QDC-triggered system.}
	\label{fig:qdc_vs_scaler}
\end{figure}

It is obvious that storing of more parameters during the measurement allows a better check of different systematic effects. However, it can also introduce a significant dead time of the acquisition system. Therefore, a two branch DAQ system was developed for WITCH (Fig.~\ref{fig:daq-scheme}). The first branch (Fig.~\ref{fig:daq-scheme}(a)) performs a simple counting and uses an  ORTEC\mbox{\scriptsize{$^{\circledR}$}} multi-channel scaler PC card (MCS-pci) with provided software. This card records the number of events as a function of time,  accepts counting rates up to 150~MHz and allows to specify a timing channel from 100~ns to 1300~s. These characteristics perfectly satisfy the WITCH requirements of $\sim$0.1~MHz count rate and a few seconds data acquisition cycle. Another highly important parameter is zero dead time between timing channels and at the end of the cycle. When the card advances from one time bin to another a ``Channel Advanced'' TTL-signal is generated which is then used to drive the retardation supply to change to the next retardation step. A direct connection between the time channels and the retardation steps is thus provided. For instance, during the $^{124}$In measurement one retardation channel corresponded to a time bin of 100~ms within one trap load. The data acquired are displayed on-line by ORTEC\mbox{\scriptsize{$^{\circledR}$}} software while they are saved to a hard-disk after 10  cycles. The second branch (called ``event-by-event'', Fig.~\ref{fig:daq-scheme}(b)) is based on a CAMAC system and LabVIEW software and stores an additional information for every event: four delay times to reconstruct the  position of the event on the MCP detector, the charge collected by the MCP, the current retardation step and the total number of events collected since the beginning of the DAQ cycle. The charge information is important because the charge distribution, or pulse-height distribution (PHD), differs for ions and $\beta$'s. For instance, comparing PHDs for the first retardation steps and for the last (stepping goes up from 0~V to 220~V) one can see (Fig.~\ref{fig:phds_vs_retard}) that in the beginning of the retardation this distribution has a Gaussian-like part which is typical for ions while at the end it is more exponential-like, typical for $\beta$'s. This provides additional proof that the main signal measured corresponds to $^{124}$In ions. Moreover, by combining the charge information with the position of an event one can select the outer rim of the detector and check that the events registered there have the PHD typical for betas and not for ions \cite{coeck07b}, thus confirming that the ion signal was well within the detector for all retardation steps and avoiding possible systematic effects on the spectrum. The total number of events since the beginning of the cycle and the current retardation step are recorded by a 100~MHz CAMAC scaler to have a reference with the first branch of the DAQ. The read-out of the second branch is triggered by an event in the QDC module (charge to digital converter). When the CAMAC system is read the QDC and TDC (time to digital converter) modules are emptied but not the CAMAC scaler that is cleared only at the beginning of the next cycle (Fig.~\ref{fig:daq-scheme}). From the information recorded one can reconstruct the measured spectrum in three ways: (1) directly from the data acquired with the first branch (MCS-pci card) (2) using the data collected in the CAMAC scaler and (3) counting how often the QDC module was triggered for every retardation step. As can be seen from Fig.~\ref{fig:124In_meas} the first and second methods correspond pretty well to each other while method (3) misses a significant portion of the counts at high count rate. By comparing the data of the 2nd and 3d methods one can deduce the dead time of the second branch of the DAQ (Fig.~\ref{fig:qdc_vs_scaler}). The average value for two data sets is  $\tau_{dead\;time}=346.6 \pm 2.0$~$\mu$s. Thus the ``event-by-event'' DAQ provides very important information to check a number of systematic effects but has a significant dead time, such that the recoil ion spectrum has to be fitted on the basis of the data collected with the first branch of the DAQ.

The usual measurement cycle starts when the ions are transferred into the decay trap. Therefore,  the transfer signal triggers the whole data acquisition (Fig.~\ref{fig:daq-scheme}). The falling edge of the transfer TTL-signal gives the start for the MCS-pci card. The width of the time channel and the number of channels are specified via the software. This number of time bins has to correspond to the number of voltage steps of the retardation supply. The start of the MCS-pci card is used to generate a ``measurement window'' to avoid further counting after the decay trap is emptied. This start signal also triggers the retardation supply so that it becomes ready for stepping. The change of the retardation step is driven by the change of the time channel in the MCS-pci card. The signal from the rear face of the MCP is amplified, delayed and sent to the QDC for the charge measurement but it is also discriminated and put in coincidence with the ``measurement window''. The NIM logic signal after the coincidence unit is then used as a QDC gate and to give a ``Common start'' for the TDC. This NIM signal is also sent for counting to the ``Event In'' input of the MCS-pci card and to the CAMAC scaler. The four signals from the delay lines of the MCP are amplified, discriminated and then sent to the stop inputs of the TDC. The whole cycle can be as well adopted for other measurement scenarios, for instance for the Off/On sequence mentioned above (sec.~\ref{spec}).

\section{Outlook}
\label{outlook}
There is still room for further developments. Here we mention several issues. The discharge problems definitively have to be investigated in more detail. Sec.~\ref{daq} clearly suggests that the ``event-by-event'' branch has to be improved in order to reduce the dead time. One of the possibilities is to use a flash ADC or fast digitizer to record the whole shape of the MCP signal and then use a software routine to analyze it. However, this will significantly increase the amount of data to be stored. The already existing ``event-by-event'' system may require up to $\sim$50~Gbyte/day. Another fast available solution is to use the existing ISOLDE DAQ system based on VME electronics and MBS software, the standard DAQ software at GSI (Germany). This system is not yet the ideal solution but it should give one order of magnitude reduction of dead time. Another ongoing development is related to a scintillator $\beta$-detector to be installed in the decay trap \cite{kozlov-tcp06}. This will allow to have a normalization between different trap loads, to perform TOF measurements, etc. The first prototype was produced in Prague and it is currently being tested in Leuven. One can also improve the DAQ cycle such that the retardation steps close to the end point of the spectrum, i.e. with less statistics, are measured longer. A last idea, but not least is to perform a case study for an energy sensitive detector instead of the currently used MCP. Ideally this will allow to separate the different charge states of the recoil ions and to distinguish ions from betas by their energy. The key issue in this separation is that the ions of different charges have different energies due to the electrostatic re-acceleration (i.e. 
$E_{ion}=q\cdot U_{acc}$). With $U_{acc}>30$~kV the ions may have enough energy, for instance to pass the dead layer of solid state detector and leave sufficient ionization signal \cite{funsten04}. This ion signal should as well be different from the signal left by $\beta$'s with typical energies of the order of 1~MeV. 

\section{Conclusion}
\label{conclusion}
The WITCH experiment opens new possibilities for recoil ion spectrometry with the primary goal to test the Standard Model for the presence of scalar and tensor exotic interactions. Recently the proof-of-principle experiment was successfully completed and the first recoil ion spectrum was obtained. Details of the measurement procedure with emphasis on the current version of DAQ system were presented. Also several checks of the validity of the data obtained were described. Further developments, in order to improve the functionality of the experiment, were briefly discussed as well.

\pdfbookmark[1]{Acknowledgements}{acknow}
\section*{Acknowledgements}
\label{acknow}
This work is supported by the European Union grants FMRX-CT97-0144
(the EUROTRAPS TMR network) and HPRI-CT-2001-50034 (the NIPNET RTD
network), by the European Union Sixth Framework through RII3-EURONS (Contract No. 506065), by the Flemish Fund for Scientific Research FWO and by the projects GOA 99-02 and GOA 2004/03 of the K.U. Leuven.



\pdfbookmark[1]{References}{refs}

\end{document}